
\magnification=\magstep1

\tolerance=10000
\voffset=-1.0cm
\baselineskip14pt
\parskip=0truemm
\voffset= -1.0truecm
\hoffset= -0.5truecm
\hsize17.0truecm
\vsize25.5truecm
\def\subtitle #1 {\vskip1.0truecm \noindent{\bf #1} \vskip0.6truecm}

\font\rmsta=cmr12

\font\bfone=cmbx12 scaled\magstep1

%
\catcode`@=11
%
%
\def\b@lank{ }

\newif\if@simboli
\newif\if@riferimenti
\newif\if@bozze

\newwrite\file@simboli
\def\simboli{
    \immediate\write16{ !!! Genera il file \jobname.SMB }
    \@simbolitrue\immediate\openout\file@simboli=\jobname.smb}

\newwrite\file@ausiliario
\def\riferimentifuturi{
    \immediate\write16{ !!! Genera il file \jobname.AUX }
    \@riferimentitrue\openin1 \jobname.aux
    \ifeof1\relax\else\closein1\relax\input\jobname.aux\fi
    \immediate\openout\file@ausiliario=\jobname.aux}

\def\bozze{\@bozzetrue}

\newcount\eq@num\global\eq@num=0
\newcount\sect@num\global\sect@num=0

\newif\if@ndoppia
\def\numerazionedoppia{\@ndoppiatrue\gdef\la@sezionecorrente{\the\sect@num}}

\def\se@indefinito#1{\expandafter\ifx\csname#1\endcsname\relax}
\def\spo@glia#1>{} 

\newif\if@primasezione
\@primasezionetrue

\def\s@ection#1\par{\immediate
    \write16{#1}\if@primasezione\global\@primasezionefalse\else\goodbreak
    \vskip\spaziosoprasez\fi\noindent
    {\bf#1}\nobreak\vskip\spaziosottosez\nobreak\noindent}
%

\def\sezpreset#1{\global\sect@num=#1
    \immediate\write16{ !!! sez-preset = #1 }   }

\def\spaziosoprasez{40pt plus 60pt}
\def\spaziosottosez{40pt}

\def\sref#1{\se@indefinito{@s@#1}\immediate\write16{ ??? \string\sref{#1}
    non definita !!!}
    \expandafter\xdef\csname @s@#1\endcsname{??}\fi\csname @s@#1\endcsname}

\def\autosez#1#2#3{
    \global\advance\sect@num by 1\if@ndoppia\global\eq@num=0\fi
    \xdef\la@sezionecorrente{\the\sect@num}
    \def\usa@getta{1}\se@indefinito{@s@#1}\def\usa@getta{2}\fi
    \expandafter\ifx\csname @s@#1\endcsname\la@sezionecorrente\def
    \usa@getta{2}\fi
    \ifodd\usa@getta\immediate\write16
      { ??? possibili riferimenti errati a \string\sref{#1} !!!}\fi
    \expandafter\xdef\csname @s@#1\endcsname{\la@sezionecorrente}
    \immediate\write16{\la@sezionecorrente. #2 #3}
    \if@simboli
      \immediate\write\file@simboli{ }\immediate\write\file@simboli{ }
      \immediate\write\file@simboli{  Sezione
                                  \la@sezionecorrente :   sref.   #1}
      \immediate\write\file@simboli{ } \fi
    \if@riferimenti
      \immediate\write\file@ausiliario{\string\expandafter\string\edef
      \string\csname\b@lank @s@#1\string\endcsname{\la@sezionecorrente}}\fi
    \noindent\if@bozze\llap{\tt#1\quad }\fi
      {\hfill{--------------------------------------------------------------%
-------------------------------------------\par\noindent}
       \hfill{\bfone CHAPTER \the\sect@num\par\noindent}
       \hfill{--------------------------------------------------------------%
-------------------------------------------\par\noindent}
       \vskip 20pt \hfill{\bfone #2}\par\noindent \hfill{\bfone #3}}
       \nobreak\vskip 80pt\nobreak}

\def\semiautosez#1#2\par{
    \gdef\la@sezionecorrente{#1}\if@ndoppia\global\eq@num=0\fi
    \if@simboli
      \immediate\write\file@simboli{ }\immediate\write\file@simboli{ }
      \immediate\write\file@simboli{  Sezione ** : sref.
          \expandafter\spo@glia\meaning\la@sezionecorrente}
      \immediate\write\file@simboli{ }\fi\noindent
       {\hfill{--------------------------------------------------------------%
-------------------------------------------\par\noindent}
       \hfill{\bfone #2\par\noindent}
       \hfill{--------------------------------------------------------------%
-------------------------------------------\par\noindent}}
       \nobreak\vskip 80pt\nobreak}


\def\eqpreset#1{\global\eq@num=#1
     \immediate\write16{ !!! eq-preset = #1 }     }

\def\eqref#1{\se@indefinito{@eq@#1}
    \immediate\write16{ ??? \string\eqref{#1} non definita !!!}
    \expandafter\xdef\csname @eq@#1\endcsname{??}
    \fi\csname @eq@#1\endcsname}

\def\eqlabel#1{\global\advance\eq@num by 1
    \if@ndoppia\xdef\il@numero{\la@sezionecorrente.\the\eq@num}
       \else\xdef\il@numero{\the\eq@num}\fi
    \def\usa@getta{1}\se@indefinito{@eq@#1}\def\usa@getta{2}\fi
    \expandafter\ifx\csname @eq@#1\endcsname\il@numero\def\usa@getta{2}\fi
    \ifodd\usa@getta\immediate\write16
       { ??? possibili riferimenti errati a \string\eqref{#1} !!!}\fi
    \expandafter\xdef\csname @eq@#1\endcsname{\il@numero}
    \if@ndoppia
       \def\usa@getta{\expandafter\spo@glia\meaning
       \la@sezionecorrente.\the\eq@num}
       \else\def\usa@getta{\the\eq@num}\fi
    \if@simboli
       \immediate\write\file@simboli{  Equazione
            \usa@getta :  eqref.   #1}\fi
    \if@riferimenti
       \immediate\write\file@ausiliario{\string\expandafter\string\edef
       \string\csname\b@lank @eq@#1\string\endcsname{\usa@getta}}\fi}

\def\autoreqno#1{\eqlabel{#1}\eqno(\csname @eq@#1\endcsname)
       \if@bozze\rlap{\tt\quad #1}\fi}
\def\autoleqno#1{\eqlabel{#1}\leqno\if@bozze\llap{\tt#1\quad}
       \fi(\csname @eq@#1\endcsname)}

\def\numeriadestra{\let\autoeqno=\autoreqno}
\def\numeriasinistra{\let\autoeqno=\autoleqno}
\numeriadestra

\newcount\cit@num\global\cit@num=0

\newwrite\file@bibliografia
\newif\if@bibliografia
\@bibliografiafalse

\def\lp@cite{[}
\def\rp@cite{]}
\def\trap@cite#1{\lp@cite #1\rp@cite}
\def\lp@bibl{[}
\def\rp@bibl{]}
\def\trap@bibl#1{\lp@bibl #1\rp@bibl}

\def\refe@renza#1{\if@bibliografia\immediate        
    \write\file@bibliografia{
    \string\item{\trap@bibl{\cref{#1}}}\string
    \bibl@ref{#1}\string\bibl@skip}\fi}

\def\ref@ridefinita#1{\if@bibliografia\immediate\write\file@bibliografia{
    \string\item{?? \trap@bibl{\cref{#1}}} ??? tentativo di ridefinire la
      citazione #1 !!! \string\bibl@skip}\fi}

\def\bibl@ref#1{\se@indefinito{@ref@#1}\immediate
    \write16{ ??? biblitem #1 indefinito !!!}\expandafter\xdef
    \csname @ref@#1\endcsname{ ??}\fi\csname @ref@#1\endcsname}

\def\c@label#1{\global\advance\cit@num by 1\xdef            
   \la@citazione{\the\cit@num}\expandafter
   \xdef\csname @c@#1\endcsname{\la@citazione}}

\def\bibl@skip{\vskip 0truept}


\def\stileincite#1#2{\global\def\lp@cite{#1}\global
    \def\rp@cite{#2}}
\def\stileinbibl#1#2{\global\def\lp@bibl{#1}\global
    \def\rp@bibl{#2}}

\def\citpreset#1{\global\cit@num=#1
    \immediate\write16{ !!! cit-preset = #1 }    }

\def\autobibliografia{\global\@bibliografiatrue\immediate
    \write16{ !!! Genera il file \jobname.BIB}\immediate
    \openout\file@bibliografia=\jobname.bib}

\def\cref#1{\se@indefinito                  
   {@c@#1}\c@label{#1}\refe@renza{#1}\fi\csname @c@#1\endcsname}

\def\cite#1{\trap@cite{\cref{#1}}}                               
\def\ccite#1#2{\trap@cite{\cref{#1},\cref{#2}}}                  
\def\cccite#1#2#3{\trap@cite{\cref{#1},\cref{#2},\cref{#3}}}     
\def\ncite#1#2{\trap@cite{\cref{#1}--\cref{#2}}}        
\def\upcite#1{$^{\,\trap@cite{\cref{#1}}}$}           		 
\def\upccite#1#2{$^{\,\trap@cite{\cref{#1},\cref{#2}}}$}  	 
\def\upcccite#1#2#3{$^{\,\trap@cite{\cref{#1},\cref{#2},\cref{#3}}}$} %
\def\upncite#1#2{$^{\,\trap@cite{\cref{#1}-\cref{#2}}}$}  

\def\clabel#1{\se@indefinito{@c@#1}\c@label           
    {#1}\refe@renza{#1}\else\c@label{#1}\ref@ridefinita{#1}\fi}

\def\biblskip#1{\def\bibl@skip{\vskip #1}}           

\def\insertbibliografia{\if@bibliografia             
    \immediate\write\file@bibliografia{ }
    \immediate\closeout\file@bibliografia
    \catcode`@=11\input\jobname.bib\catcode`@=12\fi}


\def\commento#1{\relax}
\def\biblitem#1#2\par{\expandafter\xdef\csname @ref@#1\endcsname{#2}}


\catcode`@=12

\autobibliografia
\centerline{\bfone Solid--on--solid Model for Adsorption on Self--Affine}
\centerline{\bfone Substrate: a Transfer Matrix Approach}
\vskip1.0truecm
\centerline{\rmsta G. Giugliarelli}
\centerline{\it Dipartimento di Fisica, Universit\`a di Udine, Udine, Italy}
\medskip
\centerline{and}
\centerline{\rmsta A. L. Stella}
\centerline{\it Dipartimento di Fisica e Sezione INFN, Universit\`a di
Padova, Padova, Italy}
\vskip2.0truecm
\centerline{\bf Abstract}
\bigskip
{\hsize=15truecm
 \baselineskip=12pt
 \parindent=2.0truecm

\item{} We study a $d=2$ discrete solid--on--solid model of complete wetting
        of a rough substrate with random self--affine boundary, having
        roughness exponent $\zeta_s$. A suitable transfer matrix approach
        allows to discuss adsorption isotherms, as well as geometrical and
        thermal fluctuations of the interface. For $\zeta_s\leq 1/2$ the same
        wetting exponent $\psi=1/3$ as for  flat substrate is obtained for
        the dependence of the coverage, $\theta$, on the chemical potential,
        $h$ ($\theta\sim h^{-\psi}$ for $h\to 0$). The expected existence of
        a zero temperature fixed point, leading to $\psi=\zeta_s /(2-\zeta_s)$
        for $\zeta_s>1/2$, is verified numerically in spite of an unexpected,
        very slow convergence to asymptotics.

\medskip\noindent
\item{} PACS numbers: 68.45, 68.55.Jk, 05.40

}

\vskip2.0truecm
\centerline{(Submitted for publication to PHYSICA A)}

\vfill{Email: giugliarelli@udphvx.fisica.uniud.it,
gilberto@udibm1.fisica.uniud.it}

\vfill\eject
\subtitle{1. Introduction }

   Adsorption of a nearly saturated vapour on a rough substrate has
both fundamental and technological relevance. In the case of complete
wetting \cite{dietrich/a}, which is the main concern of the present paper,
one considers the approach to infinity of the thickness of
the adsorbed liquid film in presence of the vapour phase, as the
chemical potential tends to the coexistence value. Indicating by $l$
such thickness and by $h$ the chemical potential difference, one
expects $ l \propto h^{-\psi}$.  Once assumed that roughness has
well--defined scaling properties, globally characterized by a fractal
dimension or a roughness exponent, the power law describing complete
wetting should depend on these dimensions
\clabel{degennes}\clabel{pfeifer/a}\clabel{kardar/a}\clabel{kardar/b}
\clabel{pfeifer/b}\clabel{andelman}\clabel{giugliarelli}\clabel{forgacs}
\ncite{degennes}{forgacs}.
For example, a
self--affine surface is characterized by a roughness exponent
$\zeta_s$, which gives the average transverse height fluctuation,
$\overline{|\Delta S|}$, sampled on a portion of the surface with
projection of linear size
$L$, as $\overline{|\Delta S|} \propto  L^{\zeta_s}$.\footnote{*}{Here and
in the following the overbar indicates average over disorder}
The full
elucidation of the scaling behaviours describing wetting and/or the
crossover phenomena possibly affecting their detectability, is a hard
problem, whose solution should also suggest useful new ways of
probing roughness in experimental samples.

   In recent years there have been many attempts to solve such issues,
mostly on the basis of relatively simple phenomenological models and
scaling considerations \ncite{degennes}{andelman}. Scaling laws describing
complete wetting have
been proposed for both fractal \ccite{degennes}{pfeifer/a} and self--affine
substrates \ncite{kardar/a}{andelman}.
Attempts to apply such results to the analysis of recent experiments
concerning the adsorption of N$_2$ on deposited Ag
\ccite{pfeifer/a}{chiarello} or cyclohexane vapor
on Si substrates \cite{tidswell} have also been made. Many
ambiguities however remain in this respect, because, even the recognition
of a fractal or self--affine character of the samples seems to be sometimes
a matter of controversy \ncite{pfeifer/a}{kardar/b}\cite{chiarello}.
An important step towards a satisfactory
theoretical control of wetting on rough substrates should be made by
studying in detail statistical models incorporating all essential features
of the problem. Being free of the ambiguities still affecting experiments
interpretation, such a study would provide crucial tests of the existing
conjectures and of their ranges of validity.

  In the case of fractal substrates a model investigation of this
kind has been undertaken recently by the present authors
\cite{giugliarelli}. Due to
the strong limitations imposed by the presence of overhangs for a
fractal boundary, the statistical treatment had to be limited to a
mean field approximation. However, neglecting interface fluctuations
is not expected to lead to substantial modifications of the asymptotic
results in that case. Indeed, as we also confirm below on the basis
of our results, wetting of an extremely rough surface, like a fractal,
can be considered, in comparison to a self--affine one, as a $T=0$
phenomenon \cite{lipowsky}. On the other hand,
thermal effects are still adequately described in a mean field
treatment of non asymptotic features of adsorption isotherms,
such as layering and capillary condensation phenomena. However, the
computational complexity inherent even in such mean field description,
limits considerably the possibility of detecting asymptotic laws,
like the expected power law dependence of the coverage on chemical
potential \cite{giugliarelli}.

  Self--affine substrates can be modelled by profiles without
overhangs. Due to this simplification, they are relatively more
accessible to statistical treatment. At least for moderate self--affine
roughness, thermal fluctuations of the interface are expected to play
a key role in determining wetting exponents, as it is the case for flat
substrates when the bulk dimension is lower than $3$ \cite{lipowsky}. Thus, an
adequate treatment of self--affine substrates must describe correctly
interface fluctuation phenomena. In the present paper we
accomplish this task by a transfer matrix study of an interface model of
adsorption on self--affine boundary in $2$ bulk dimensions. The
interface is treated in a solid--on--solid (SOS) approximation and
takes the shapes of a directed self--avoiding walk on square
lattice. Thermal interface fluctuations are controlled by surface tension and
lead to self--affine behavior with a roughness exponent $\zeta=1/2$
\cite{forgacs}.  The
boundary of the substrate has the same kind of shapes, with quenched
fluctuations. Generalization of a method introduced by Mandelbrot
\cite{mandelbrot}
enables us to produce random boundaries with preassigned roughness
exponents, $0<\zeta_s<1$. The liquid in the wetting film is further
supposed to feel a van der Waals type of potential from the
substrate, and its thickness $l$ is controlled by the chemical potential.

   Existing scaling and analytical approaches to this problem
\ccite{kardar/b}{li} have strong ties with the work of Fisher and Lipowsky
\cite{lipowsky}, who first discussed the role of the interfacial $\zeta$ in
determining exponents like $\psi$ in the flat case. The basic idea emerging
from such scaling arguments, which will be recalled in the next section, is
that for complete wetting the asymptotics should be determined by the
maximum between $\zeta$ and $\zeta_s$. When $\zeta>\zeta_s$, the $\psi$
exponent is the same as in the flat case, and thermal fluctuations control
the asymptotics. When $\zeta<\zeta_s$, control is assumed by the quenched
geometrical fluctuations, and the role of $\zeta$ is taken by $\zeta_s$ in
the formulas. Even if replica based perturbative renormalization group
calculations have confirmed this simple scheme \cite{li}, and the fact that
only $\zeta_s$ determines scaling in the strong fluctuation regime, with
$\zeta_s>\zeta$, a direct verification did not exist and is one of the
goals of our study.

   The application of transfer matrix techniques to a model like the one
studied here encounters some complications which we
explain below, and involves relatively heavy computing. Thus, the
approach presented in this paper has some methodological interest and should
also constitute a starting point in view of further applications.

   This paper is organized as follows. In the next section we
introduce the model, discuss some technicalities of the transfer
matrix approach, and present a simple derivation of the scaling
results which are the object of our numerical investigation. Section
$3$ is devoted to a discussion of test applications to the case of flat
substrate. In section $4$ we present the main results for substrates with
different degrees of roughness, with particular emphasis on the basic scaling
laws involved. The last section contains general conclusions and outlines
some open perspectives.

\subtitle{2. SOS Model, Transfer Matrix Approach and Scaling Arguments}

   In the SOS approximation the separation between condensed and vapour
phases of the wetting fluid is described by a sharp interface, which
fluctuates without forming overhangs. In 2 bulk dimensions on a square
lattice this interface can be represented by a single valued integer
function $Z_x$ of the abscissa $x$, and fluctuations are controlled by a
Hamiltonian of the form:
$$
{\cal H}=\sum_x \left[{\varepsilon\over 2}|Z_x-Z_{x+1}| + U_x(Z_x)+hZ_x
\right]
\eqno(1)
$$
where $x$ also runs over integers. The first term in the sum of eq.(1)
represents the energy cost for the interface steps due to surface tension.
Indeed, $\varepsilon$ corresponds to the strength of the attractive
nearest--neighbor energy between fluid particles in a lattice--gas model,
in which particles can be placed at the centers of each elementary
square not already occupied by substrate atoms.
The second term in the sum is the energy due to long range van der Waals
interaction with the substrate and $h$ is the deviation of chemical
potential from the coexistence value.

   The substrate boundary can be represented by a single valued function
$S_x$. Upon averaging over randomness, the following scaling is expected to
hold
$$
\overline{|S_x-S_{x'}|}\sim |x-x'|^{\zeta_s},
\eqno(2)
$$
$\zeta_s$ being the surface roughness exponent, with values in the
interval $(0,1)$.

   The interface heights can be written as $Z_x=S_x+z_x$ where $z_x$
is the local thickness of the wetting film, i.e. the distance
of the SOS interface from the substrate.

   $U_x(Z_x)$ in eq.(1) represents the potential energy, due to the
substrate, of a column of fluid with height $z_x$ at $x$. The long range
nature of the potential and numerical checks convinced us that the following
local approximation for this function
$$
U_x(Z_x)= -U_x^{(0)}+c/z^{\sigma-1},
\eqno(3)
$$
is adequate to reproduce the scaling properties of adsorption in the
high coverage regime. In all calculations reported here the choice
$\sigma=4$ has been made. This corresponds to a $1/r^6$ van der Waals
interparticle potential.

   We thus rewrite Hamiltonian (1) in the form:
$$
{\cal H}= -\sum_z U_x^{(0)}+\sum_x \left[{\varepsilon \over 2}
|z_x-z_{x+1}+\Delta S_x| +{c \over {z_x^{\sigma-1}}}+hz_x\right]
\eqno(4)
$$
where $\Delta S_x=S_x-S_{x+1}$ are the substrate surface steps and $L$
is the horizontal linear size of the system.
Here and in the following we assume $1\leq z <\infty$.

   We have set up a transfer matrix technique \ccite{forgacs}{privman}
for the solution of the
problem. In our case, because of roughness, and according to Hamiltonian
(4), we can associate to any $x$ two transfer matrices, ${\bf T}_+(x)$ and
${\bf T}_-(x)$, with elements
$$
[{\bf T}_+(x)]_{z,z'} = \exp \left[-({\varepsilon
\over 2} |z-z'+\Delta S_x| + {c \over {z^{\sigma-1}}}+hz)/T \right]
\eqno(5a)
$$
$$
[{\bf T}_-(x)]_{z,z'} = \exp \left[-({\varepsilon \over 2}
|z-z'-\Delta S_{x-1}| + {c \over {z^{\sigma-1}}}+hz)/T \right]
\eqno(5b)
$$
where $T$ is the temperature measured in units of energy. The first
term in eq.(4) does not enter in the definition of (5$a$--$b$) since it does
not affect adsorption properties. One can see that $[{\bf T}_+(x)]_{z,z'}$
and $[{\bf T}_-(x)]_{z,z'}$ represent the Boltzman weighing factor for a
step of the SOS interface from $(x,z)$ to $(x+1,z')$ or $(x-1,z')$,
respectively.

   A film of adsorbed liquid can be represented by a very long SOS
interface with,e.g., its ends pinned by the surface (see Fig.1). The average
distance of the interface from the substrate calculated in its middle
point will give a measure of the thickness of the wetting film.
The evaluation, at a given temperature, of such thickness
as a function of chemical potential $h$ gives the adsorption isotherms.

   Let us consider a SOS interface of length $L$. This interface can ideally
be separated into two parts at the middle point
$x_m$: for example, the left part has the left--end pinned at $(x_m-L/2,1)$
and the right--end at $(x_m,z)$ (see Fig.1). We begin by treating the two
parts as
mutually independent. Let us consider first the left one.
While its left--end is pinned, as we
move to the right, the interface becomes more and more free to
fluctuate. The position of the interface at $x_m$ can be
studied in terms of an occupation profile $\eta_+(x_m,z)$ which is
the probability for the interface to be at $(x_m,z)$.
It is easy to verify that
$$
\eta_+ (x_m,z) \propto [{\bf T}_+(x_m-L/2)\cdot {\bf T}_+(x_m-L/2+1)
\cdots {\bf T}_+(x_m-1)]_{1,z}.
\eqno(6)
$$
This follows from iterating the formula:
$$
\eta_+(x+1,z)=N_{x+1}^+\sum_{z'} \eta_+(x,z')[{\bf T}_+(x)]_{z',z}
\eqno(7)
$$
where $N_{x+1}^+=1/\sum_{z',z''} \eta(x,z')[{\bf T}_+(x)]_{z',z''}$ is a
normalization constant ($\sum_z \eta_+(x,z)\equiv 1$) and
$\eta_+(x_m-L/2,z)\equiv \delta_{z,1}$.
In general, the shape of the occupation profile in $x_m$ will depend not
only on $L$ but also on the shape of the substrate surface between
$x_m-L/2$ and $x_m$. However, we expect that for large $L$
the interface will loose memory of the starting
point, and $\eta_+(x_m,z)$ will depend only on the shape of the
substrate within a distance $\xi_\parallel$, to be defined
below.

   For the right part of the interface we can obviously repeat similar
considerations. The occupation profile in this case will be obtained by
iterations of the form:
$$
\eta_-(x-1,z)=N_{x-1}^-\sum_{z'} \eta_-(x,z')[{\bf T}_-(x)]_{z',z}
\eqno(8)
$$
The occupation profiles at $x_m$ we can get from eqs. (7) and (8)
will in general differ from each other (see Fig.1). This is due to
the roughness of the substrate which makes the surface asymmetric.

   In order to determine a unique, continuous interface on the basis
of $\eta_+$ and $\eta_-$, it is natural to impose that the left and
right interfaces match at $x_m$. We thus define a profile:
$$
\eta(x_m,z)=N_{x_m} \eta_+(x_m,z) \eta_-(x_m,z)
\eqno(9)
$$
where $N_{x_m}=1/\sum_z \eta_+(x_m,z) \eta_-(x_m,z)$ is the corresponding
normalization factor. Of course, the same procedure discussed for $x_m$ and
leading to eq.(9) can be applied to any $x$, leading to the construction of
a unique profile, as illustrated in Fig. 1.

   To obtain macroscopic quantities, two average operations have to be
performed. We will distinguish between thermal and disorder averages,
marked by brakets and overbars, respectively. The main quantity we are
interested in is the coverage, i.e. the thickness of the wetting film,
$\theta$. $\theta$ is given by
$$
\theta\equiv \overline{\langle z(x) \rangle}=\overline{\sum_z z\eta(x,z)}
            =\sum_z z\ \overline{\eta(x,z)}
\eqno(10)
$$

   On the other hand, also interface roughness has to be considered.
In the case of a flat substrate, interface roughness is due to
thermal fluctuations alone. Here, we can not exclude that the
interface roughness can also be affected by the self--affine
substrate. Information on the interface is given by
the difference correlation function
$$
\Delta C(x-x') \equiv {1 \over 2}\overline{\langle [z(x)-z(x')]^2
                  \rangle} = \overline{\langle z^2(x)\rangle}-
          \overline{\langle z(x)z(x')\rangle}
\eqno(11)
$$
The last thermal average on the r.h.s of eq.(11) can be defined
following lines similar to those discussed above for the case of a
quantity depending on one $x$ alone. An idea of how this can be
achieved is given in the following section, where an application to
the flat case is discussed.
The roughness of the interface can be measured by the perpendicular
correlation length $\xi_\perp$ which is extracted from $\Delta C$
in the $|x-x'|\to\infty$ limit
$$
\xi_\perp^2=\Delta C(\infty)=\overline{\langle z^2(x)\rangle}-
          \overline{\langle z(x)\rangle}^2
\eqno(12)
$$
It is interesting to note that, by summing and subtracting the
quantity $\overline{\langle z(x)\rangle^2}$ to eq.(12),
$\xi_\perp$ can be expressed in the form
$$
\xi_\perp^2=\xi_{\perp,T}^2+\xi_{\perp,R}^2
\eqno(13)
$$
with
$$
\xi_{\perp,T}^2 = \overline{\langle z^2(x)\rangle}-
 \overline{\langle z(x)\rangle^2}=\overline{\langle z^2(x)\rangle-\langle
  z(x)\rangle^2}
\eqno(14a)
$$
and
$$
\xi_{\perp,R}^2 = \overline{\langle z(x)\rangle^2}-
                   \overline{\langle z(x)\rangle}^2
\eqno(14b)
$$
Thus the width of the interface is
the sum of a thermal term ($\xi_{\perp,T}$), which is just the
quenched average of the local mean square fluctuation around the mean
position
of the interface, and a geometrical one, i.e. the mean square fluctuation,
induced by the surface geometrical disorder, of the average mean position
($\xi_{\perp,R}$). The second term
clearly vanishes when a flat surface is considered.

   The parallel correlation length $\xi_\parallel$ can also be calculated
from $\overline{\langle z(x)z(x')\rangle}$. Indeed one expects
\cite{forgacs}
$$
\Delta C(\infty) - \Delta C(x-x')\sim \exp(-|x-x'|/\xi_\parallel)
\eqno(15)
$$
in the $|x-x'|\to\infty$ limit.

   As we verify in the next section, for an interface bound near a
flat substrate,
$$
\xi_\perp\sim \xi_\parallel^\zeta
\eqno(16)
$$
with $\zeta=1/2$ in $d=2$. This interface wandering exponent in $d=2$
is explained by the fact that a free interface performs a random walk
in the vertical direction \cite{forgacs}.

   Following Lipowsky and Fisher \cite{lipowsky}, we can separate into two
terms
the free energy of a wetting film of average depth $l$. The first one
is $U(l)=hl+ c/l^{\sigma-1} +const.$, and represents the free energy
contribution due to the last two terms of the second sum in eq.(4).
The second term can be written on the basis of the standard continuum
interface Hamiltonian, which replaces the first term in the above
mentioned sum by a gradient squared of the local depth. For our
interface, this gradient can be represented by
$\xi_\perp/\xi_\parallel\sim \xi_\perp^{(\zeta-1)/\zeta}$. Moreover,
since for a bound interface it is natural to expect $l\approx \xi_\perp$,
the free energy density due to interface roughness, takes the form:
$$
f_I\simeq vl^{2(\zeta-1)/\zeta}.
\eqno(17)
$$
In the case of a flat substrate, and in the absence of any kind of
randomness, for $d=2$, $\zeta=1/2$ has to be assumed in eq.(17), and
minimization of $U+f_I$ gives
$$
\eqalign{
{d\over{dl}} (U+f_I) &= h-{{(\sigma-1)c}\over{l^\sigma}}-2vl^{-3}\cr
                     &\simeq h-vl^{-3}=0\cr
                     &^{\l\to\infty}\cr
        }
\eqno(18)
$$
for $\sigma>3$, which leads to $l\simeq\theta\propto h^{-1/3}$, i.e.
$\psi=1/3$.

   If the substrate is rough, with exponent $\zeta_s$, the problem
arises to establish which $\zeta$ should actually enter is eq.(17). It
is rather natural to expect that, as long as $\zeta_s<1/2$, the
intrinsic thermal roughness of the interface determines $f_I$. Thus
$\zeta=1/2$ has to be chosen in eq.(17) and the flat case $\psi=1/3$
still holds. On the other hand, for $\zeta_s>1/2$, the geometrical
substrate roughness induces a corrugation energy for the interface,
which can still be estimated via eq.(17), this time with
$\zeta=\zeta_s>1/2$.

   This energy term becomes now asymptotically dominant in the
minimum condition:
$$
{d\over{dl}}(hl-v'l^{2(\zeta_s-1)/\zeta_s})=0
\eqno(19)
$$
yielding $\psi=\zeta_s/(2-\zeta_s)$, $\zeta_s\geq 1/2$. In eq.(19) we
again exploited the fact that $l\approx \xi_\perp$.

   These results were first derived by Kardar and Indekeu \cite{kardar/b}
on the basis
of similar scaling arguments. An important remark to be made concerns
the fact that the replacement of $\zeta=1/2$ by $\zeta=\zeta_s>1/2$ in
eq.(17), leads to an $f_I$ which violates hyperscaling. Indeed, since
$f_I$ is a free energy per unit horizontal length, it should scale
like $\xi_\parallel^{-1}\sim \xi_\perp^{-1/\zeta}$. For $\zeta=1/2$,
taking into account that $l\sim\xi_\perp$, $f_R\sim l^{-2}$ is indeed
consistent with hyperscaling. For $\zeta=\zeta_s>1/2$, the roughness
free energy in eq.(19) does not scale as $\xi_\parallel^{-1}$. The
hyperscaling violation is consistent with a zero temperature fixed
point describing complete wetting. In the regime $\zeta_s>1/2$,
interface fluctuations are of a quenched geometrical, rather than
thermal nature. A direct verification of this will be obtained in
section 4, where it will be shown that while $\xi_\perp\simeq
\xi_{\perp,T}>>\xi_{\perp,R}$ for $\zeta_s\leq 1/2$,
$\xi_\perp\simeq\xi_{\perp,R}>>\xi_{\perp,T}$ for $\zeta_s>1/2$.

\subtitle{3. Flat Substrate}

   As a preliminary test of the transfer matrix approach, we have
considered the case of a flat substrate. The absence of
geometrical disorder determines here several simplifications. In
particular, the transfer matrices ${\bf T}_+(x)$ and ${\bf T}_-(x)$
in eqs. (5$a$--$b$) are equal to each other. Thus, a unique
$x$--independent matrix, ${\bf T}$, has to be considered. As a
consequence, the occupation profiles $\eta_+$ and $\eta_-$ also
coincide, and $\eta_+(x,z)=\eta_-(x,z) \equiv\eta_0(z)$.

   One should note that the present case could be treated,
equivalently, by computing directly the partition function of the
system, i.e. ${\cal Z}=Tr[{\bf T}^L]$. For large sizes, $L$, of
the system, $\cal Z$ can be approximated by the quantity $\lambda^L$
where $\lambda$ is the largest eigenvalue of ${\bf T}$. The relation
with our approach is due to the fact that $\eta_0$ is the eigenvector
of ${\bf T}$ associated to $\lambda$ \cite{privman}.

   Thus, the above defined occupation profile $\eta$ is now given by
$\eta(z)=N \eta_0^2(z)$. In spite of the
absence of disorder, $\eta_0$ cannot be calculated analytically. So,
we evaluated it numerically by iterating eq.(7) until convergence.
The coverage $\theta$ is then obtained by eq.(10), in which obviously
disorder averaging does not apply. $\xi_\perp$ follows from eq.(12)
or (14) ($\xi_{\perp,R}\equiv 0$). The limited complexity of the
numerical calculations makes also the evaluation of the interface
parallel correlation length, $\xi_\parallel$, feasible. In fact, for
the calculation of $\overline{\langle z(x)z(x')\rangle}$, which is now
equal to $\langle z(x)z(x')\rangle$, we can use the expression
$$
\langle z(x)z(x')\rangle ={{\sum_{z,z'} z\eta(z) [{\bf
T}^{|x-x'|}]_{z,z'} \eta(z') z'}\over{\sum_{z,z'} \eta(z)[{\bf
T}^{|x-x'|}]_{z,z'} \eta(z')}},
\eqno(20)
$$
which straightforwardly follows from extending considerations
made in the previous section.

   To work in complete wetting conditions the parameters
$\varepsilon$, $c$ and $T$ have to be conveniently fixed. In
particular we have chosen the values $c/\varepsilon=2$, for
which the system is wetted at any temperature (i.e. $T_w=0$). Our
choice is meant to represent a typical situation for
complete wetting and is not far from values appropriate
for real adsorption experiments, like krypton on graphite \cite{deoliveira},
argon on solid xenon \cite{ebner}, or $N_2$ on $Ag$ \cite{pfeifer/a}.

   To test the asymptotics of adsorption, which corresponds
to consider very thick adsorbed films, the size of
${\bf T}$ should be made as large as possible. Of course, in a
numerical calculation a reasonable matrix size $z_{max}\times z_{max}$
has to be considered, realizing a
compromise between the necessity of avoiding finite size effects and
that of reducing computation times. In our runs $z_{max}$ ranged
from a minimum of 300 (low coverage) to a maximum of 800 (high
coverage), in this flat case.

   As discussed in the previous section, due to the dominance of
interfacial thermal fluctuations, adsorption isotherms should follow the
law $\theta\sim h^{-1/3}$ for $h\to 0$. On the other hand, at very low
$\theta$, the effects of the attractive potential, $U(l)$, possibly
dominate with respect to $f_I$, and a behaviour $l\simeq\theta \sim
h^{-1/\sigma}$ (see eq.(18)) could hold before the thermal fluctuation
effects are able to impose the asymptotic regime. This $\psi=1/\sigma$ is
strictly a $T=0$ exponent, of the FHH type (see Ref. \cite{frenkel}), and
the low coverage window in which it can manifest itself will be the narrower,
the higher the temperature.

   In Fig. 2$a)$ we report three isotherms at temperatures
$T/\varepsilon=0.2$ (open squares),
$T/\varepsilon=0.4$ (open circles) and $T/\varepsilon=0.8$ (heavy
circles). In the isotherm at higher $T$ the interface
fluctuations are dominating already at very low $\theta$, and the
log--log plot of $\theta$ versus $h$ (heavy  circles) is everywhere
consistent with $\psi=1/3$ (slope of the dashed line). To further
confirmation of this scaling behaviour, a fit of the isotherm with a
function of the type $A h^{-\psi}+B$ yielded
$\psi=0.327\pm0.004$. For this case, the plot in the insert shows
also that $\xi_\perp$ follows the same scaling law and essentially
$\xi_\perp \simeq \theta$ which is
consistent with the assumptions made at the end of the previous
sections. In Fig. 2$b)$, for the two cases at higher temperatures reported in
Fig.2$a)$, the
corresponding  $\xi_\perp$ are plotted versus the respective
$\xi_\parallel$ obtained by eq.(20). The log--log plot shows that
eq.(16) is obeyed with $\zeta=1/2$ (dashed line slope), as
expected. Fitting of the data for the higher temperature case, with a
function of
the type $A{\xi_\parallel}^\nu$ resulted in a $\nu=0.478\pm0.002$.
The two isotherms represented by open circles and squares in Fig. 2$a)$
demonstrate that, the lower is the temperature, the wider is the
preasymptotic region of $\theta$ for which
the scaling behaviour is not
far from the law $\theta\sim h^{-1/\sigma}$ ($\sigma=4$ in the actual
example). The above results show that, at least in the case of flat
substrate, complete wetting is correctly described by our
transfer matrix approach.

\subtitle{4. Complete Wetting of Self--Affine Surfaces}

   As mentioned in the introduction, our self-affine boundaries are obtained
by a random generalization of an algorithm introduced by Mandelbrot
\cite{mandelbrot}. This consists in applying a
recursive transformation to an initially staircase shaped lattice walk. In
practice, given an even number $n$, we consider a starting directed walk of
$2n^k$ steps, obtained by alternating elementary forward and upward steps
on a
2--$d$ square lattice. Then, given a second even number $m$, with $m<n$,
the walk is divided in $n$ equal parts and in $(n-m)/2$ of them,
chosen at random, all the vertical
steps are reversed. The same procedure is then applied to each of the $n$
parts of the walk obtained after the first stage, and all proceeds for $k$
times. It is
easy to check that the profile obtained in this way has a roughness
exponent $\zeta_s=\ln m/\ln n$. We checked numerically that the
average height--height correlation of relatively small samples of
boundary profiles obtained as above satisfied the scaling law (2), with the
expected $\zeta_s$ up to a percent, or so.

   Since our self--affine boundaries are random, quenched averaging is
needed in our evaluation of the wetting film properties. In practice we
were careful to use in each actual calculation big enough $L$'s, so as to
guarantee a high degree of self--averaging in each individual realization
of the random profile. Another technical difference, compared to the flat
case, concerns the
size of the transfer matrices. Due to the higher coverage induced by
surface roughness, a larger $z_{max}$ has to be considered. Of course,
$z_{max}$ increases with increasing $\zeta_s$. The results we obtained
required to consider matrices with $z_{max}$ between 1000 and 1500, with
lengths $L$ up to $6\cdot 10^4$.

   To test the effects of surface roughness on adsorption properties, we
have considered three values of $\zeta_s$, $\zeta_s=1/4,1/2$ and $3/4$.
According to the scaling arguments of the previous section, the first and
third values are representative of the low and high roughness regimes,
respectively. $\zeta_s=1/2$ marks the border line between the two regimes.
For all $\zeta_s$, occupation profiles were obtained by eqs. (7--9). The
coverage $\theta$ and $\xi_\perp$, $\xi_{\perp,T}$ and $\xi_{\perp,R}$ were
also computed, and, for $\theta$, the asymptotic scaling exponent $\psi$
was extrapolated. Computing time limitations did not allow a direct
evaluation of $\xi_\parallel$, at variance with the flat case. The choice
$c/\varepsilon=2$ discussed in the previous section was made in all cases.

   Fig. 3 reports the results for $\zeta_s=1/4$ boundaries. Temperature was
fixed at $T/\varepsilon=0.8$. The coverage $\theta$ (heavy circles) follows
in the whole range of $h$ values (about 6 decades), a scaling
behaviour consistent with $\psi=1/3$ (see dashed line). As a confirmation,
a fit encompassing the whole data range with a function of the
form $Ah^{-\psi}+B$ gave $\psi=0.319\pm 0.008$. Heavy squares represent
data for $\xi_\perp$, which clearly follows the same behaviour as $\theta$
($\theta\sim \xi_\perp$). In addition, the thermal part $\xi_{\perp,T}$
(open squares) gives the entire $\xi_\perp$, the geometrical one being much
lower (open rhombs). Thus, as expected, interface roughness is dominated by
thermal fluctuations.

   Fig. 4 shows the results for the case of boundaries with $\zeta_s=1/2$,
again at $T/\varepsilon=0.8$. In spite of the higher roughness of the
substrate, $\theta$ (heavy circles) and $\xi_\perp$ (heavy squares) still
follow the same scaling as for flat substrate. A fit of the type described
above gave $\psi=0.327\pm 0.002$, still very close to the expected
$\psi=1/3$. In this case $\xi_{\perp,T}$ and $\xi_{\perp,R}$ follow
essentially the same scaling law, but, still,
$\xi_{\perp,R}<<\xi_{\perp,T}$.

   In the case of high substrate roughness, it turns out that the
asymptotic scaling law expected on the basis of the $T=0$ fixed point
mentioned in the second section, does not show up easily, even at
prohibitively high coverages. For high coverages thermal fluctuations are
still very important and seem to induce a behaviour close to the $\psi=1/3$
law. In order to reduce the role of thermal fluctuations and to obtain
evidence of a crossover to the expected $\psi=\zeta_s /(2-\zeta_s)$, we had
to reduce the temperature to $T/\varepsilon=0.5$ this time, and to consider
$L=6\cdot 10^4$ and $z_{max}=1500$. With such choices, somehow at the limit
of our computational possibilities, one detects a sort of saturation of the
coverage (heavy circles in Fig. 5) at very low $h$. This saturation is due
to truncation of the transfer matrices. On the other hand, our numerical
checks indicate that truncation effects are essentially absent up to $h\geq
6 \cdot 10^{-7}$. An interesting feature of the isotherm is that, for $h$
just above this limit, the slope has a clear increase, and, for about one
decade, seems to be consistent with a $\psi=0.632\pm 0.021$. This estimate
is reasonably close to the value $\psi=\zeta_s /(2-\zeta_s)=0.6$ implied by
the scaling arguments. We interpret these results as an indication that,
after a preasymptotic regime of lower effective $\psi$, the system tends to
follow the expected law at very high coverages. In the preasymptotic regime
the slope of the log-log plot of $\theta$ is definitely lower, and rather
close to the flat case one (dashed line).

Fig. 5 also shows that
$\xi_\perp$ (heavy squares) follows the same law as $\theta$. Now, however,
interface roughness is almost completely due to geometry, rather than
temperature, and the roles of $\xi_{\perp,T}$ and $\xi_{\perp,R}$ are
interchanged with respect to the previous cases. $\xi_{\perp,T}$ (open
squares) is much lower than $\xi_{\perp,R}$ (open rhombs). It also turns out
that, in the preasymptotic regime,
$\xi_{\perp,T}$  obeys rather closely the $h^{-1/3}$ law, as in the
flat substrate case (dashed line). As $\theta$ reaches its expected asymptotic
slope, also $\xi_{\perp,R}$ rises more rapidly, and, before saturation,
assumes a slope comparable
with that of $\theta$, as well.

\subtitle{5. Concluding Remarks}

   In this paper we performed a systematic study of the effects of
self--affine substrate roughness on complete wetting properties of a
fluid. This study was based on a discrete, $d=2$ statistical
model genuinely incorporating the essential features of the problem, namely
interface fluctuations, long range substrate potential, and geometrical
disorder of the surface.

   Up to now, existing studies of complete wetting on self--affine
substrates all relied on essentially continuum descriptions, so that the
problem of testing their scaling \ccite{kardar/a}{kardar/b}
on discrete models remained open, together with that of
precisely establishing the ranges of validity of various regimes and of
locating their crossovers.

   We applied to our model a transfer matrix technique
which allowed to make a useful distinction between thermally and geometrically
driven interface fluctuations, and to relate them to the coverage properties.

   With a rather high level of accuracy we could verify that power law
behaviours expected on the basis of scaling arguments are verified already
at relatively low coverages, and for different temperature choices, in the
case of moderately rough substrates ($\zeta_s\leq 1/2$). The case
$\zeta_s=3/4$ was expected to constitute an example of complete wetting
exclusively controlled by geometrical disorder through a $T=0$ fixed point.
As a matter of fact, a satisfactory treatment of this relatively high
substrate roughness revealed a real computational challenge within our model.
We could verify that with $\zeta_s=3/4$ a very wide preasymptotic scaling
regime with $\psi$ slightly larger than $1/3$ exists. Only by suitably
reducing the effects
of thermal fluctuations, responsible for the preasymptotic behaviour, it
becomes possible to detect crossover to the expected $\psi=0.6$, with an
effort reaching almost the limits of our computational capabilities. Such
slow set up of the asymptotic regime indicates the existence of important
scaling corrections, and
was not {\sl a priori} expected. This
should constitute a warning with respect to attempts to interpret  numerical
or experimental results in the field.

   The proof that a $T=0$ fixed point indeed controls asymptotics in the
case of high roughness is an indirect confirmation that studies of complete
wetting on fractally rough substrates can be correct in predicting the
asymptotics, even if disregarding thermal fluctuations of the interface.
This applies to our previous study of adsorption on a fractal substrate by
mean field methods \cite{giugliarelli}.

   The problem of complete wetting does not at all exhaust the range of
applicability of methods like those we applied in this paper. The natural
further step, already undertaken by us, is the study of the critical
wetting transition \cccite{dietrich/a}{forgacs}{dietrich/b}, possibly
occurring when in models of the kind
discussed here the temperature approaches from below some critical value,
$T_w$.
The role of self--affine roughness in this case can not be discussed in
terms of simple scaling ideas, and more sophisticated, albeit approximated,
renormalization group methods have been produced to elucidate it \cite{li}.
In this
context the role of our model studies will reveal even more crucial, also
because, we anticipate, surface roughness can lead to new
features of wetting, at low temperature, which do not seem to be easily
catched by continuum descriptions.

\subtitle{Acknowledgements}

   The work has been supported by MURST also through
INFM, Unit\'a di Padova and by italian CNR via the Cray project of
Statistical Mechanics.


\biblitem{dietrich/a}
For a review, see S. Dietrich, in ``{\sl Phase Transition and
Critical Phenomena}'', Vol. 12, by C. Domb and J. L. Lebowitz,
Academic Press, 1988

\biblitem{pfeifer/a}
P. Pfeifer, Y. J. Wu, M. W. Cole and J. Krim, Phys. Rev. Lett.
62 (1989) 1997

\biblitem{kardar/a}
M. Kardar and J. O. Indekeu, Phys. Rev. Lett. 65 (1990) 662

\biblitem{kardar/b}
M. Kardar and J. O. Indekeu, Europhys. Lett. 12 (1990) 161

\biblitem{pfeifer/b}
P. Pfeifer and M. W. Cole, New J. Chem. 14 (1990) 221

\biblitem{andelman}
D. Andelman, J.--F. Joanny and M. O. Robbins, Europhys. Lett. 7 (1988) 731;
M. O. Robbins, D. Andelman and J.--F. Joanny, Phys. Rev. A 43
(1991) 4344

\biblitem{giugliarelli}
G. Giugliarelli and A. L. Stella, Physica Scripta T35 (1991) 34

\biblitem{forgacs}
G. Forgacs, R. Lipowsky and Th. M. Nieuwenhuizen, in
``{\sl Phase Transitions and Critical Phenomena}'',
by C. Domb and J.L. Lebowitz, Vol. 14, Academic Press, 1991

\biblitem{chiarello}
R. Chiarello, V. Panella, J. Krim and C. Thompson, Phys. Rev. Lett.
67 (1991) 3408

\biblitem{tidswell}
I. M. Tidswell, T. A. Rabedeau, P. S. Pershan and S. D. Kosowsky,
Phys. Rev. Lett. 66 (1991) 2108

\biblitem{lipowsky}
R. Lipowsky and M.E. Fisher, Phys. Rev. Lett. 56 (1986) 472

\biblitem{mandelbrot}
B. B. Mandelbrot, Physica Scripta 32 (1985) 257

\biblitem{li}
H. Li and M. Kardar, Phys. Rev. B 42 (1990) 6546

\biblitem{privman}
V. Privman and N. M. Svraki\'c, in {\sl Directed Models of
Polymers, Interfaces and Clusters: Scaling and Finite--Size
Properties}, Lecture Notes in Physics, Vol.338 (Springer--Verlag,
Berlin, 1989).

\biblitem{deoliveira}
M.de Oliveira and R.B.Griffiths, Surface Science 71 (1978) 687

\biblitem{ebner}
C. Ebner, Phys. Rev. A 22 (1980) 2776

\biblitem{frenkel}
J. Frenkel, ``{\sl Kinetic Theory of Liquids}'' Oxford
University Press, London and New York, 1946

\biblitem{dietrich/b}
S. Dietrich and M. Schick, Phys. Rev. B 31 (1985) 4718

\biblitem{degennes}
P. G. de Gennes, in ``{\sl Physics of Disordered Materials}" (D. Adler et
al. eds.), Plenum Press, New York, 1985


\vfill\eject
{\parskip=0.2truecm
\noindent {\bf REFERENCES}
\bigskip
\insertbibliografia  }

\vfill\eject
{\parskip=0.4truecm
\parindent=1.3truecm
\noindent {\bf FIGURE CAPTIONS}
\bigskip
\item{Fig. 1} Shape of a SOS interface on a
rough self--affine surface with $\zeta_s=1/2$ (represented by the stepped
curve bounding the shaded area). The heavy continuous line represents
the mean shape of the interface with the ends pinned at $A$ and $B$ on
the substrate and obtained by applying at each $x$ a matching condition of
the form (9). Dashed and long--dashed lines represent the mean shapes of
the right and left interfaces discussed in the text. In the upper plot, the
occupation profiles $\eta$, $\eta_+$, $\eta_-$ at the middle point (see
arrow) are shown. The calculation were performed with $c/\varepsilon=2$,
$T/\varepsilon=0.5$ and $h=2.44\cdot 10^{-4}$.

\item{Fig. 2} Adsorption on a flat surface.
Calculations were done with $c/\varepsilon=2$.
$a)$ Adsorption isotherms for $T/\varepsilon=0.2$ (open squares),
$T/\varepsilon=0.4$ (open circles) and
$T/\varepsilon=0.8$ (heavy circles). Dashed and long dashed lines express
a $h^{-1/3}$ and $h^{-1/4}$ scaling laws, respectively. In the insert
$\xi_\perp$ (heavy squares) for $T/\varepsilon=0.8$ is plotted versus $h$.
Also in this case
the dashed line expresses a $h^{-1/3}$ scaling law. $b)$ $\xi_\perp$
versus $\xi_\parallel$ for the two cases at higher temperature reported in
$a)$. Dashed line expresses a $\xi_\parallel^{1/2}$ scaling law.

\item{Fig. 3} Adsorption on a rough self--affine surface with
$\zeta_s=1/4$. The calculation were done with $c/\varepsilon=2$
and $T/\varepsilon=0.8$. Heavy circles and squares represent the value of
coverage $\theta$ and $\xi_\perp$ as functions of $h$, respectively.
Open squares (partially hidden by heavy squares) and open rhombs
represent $\xi_{\perp,T}$ and $\xi_{\perp,R}$ values, respectively.
The dashed line obeys the $h^{-1/3}$ scaling law.

\item{Fig. 4} Adsorption on a rough
self--affine surface with $\zeta_s=1/2$. The choice $c/\varepsilon=2$
and $T/\varepsilon=0.8$ was made in this case. Heavy circles and squares
represent
$\theta$ and $\xi_\perp$ as function of $h$, respectively.
Open squares (partially hidden by heavy squares) and open rhombs
represent $\xi_{\perp,T}$ and $\xi_{\perp,R}$, respectively.
The dashed line gives again the a $h^{-1/3}$ power law.

\item{Fig. 5} Adsorption on a rough
self--affine surface with
$\zeta_s=3/4$. The calculation were done with $c/\varepsilon=0.5$
and $T/\varepsilon=0.5$. Heavy circles and squares represent
$\theta$ and $\xi_\perp$ as functions of $h$, respectively.
Open squares and open rhombs
give $\xi_{\perp,T}$ and $\xi_{\perp,R}$ values, respectively.
The solid line results from fitting $\theta$ with a function of the form
$Ah^{-\psi}+B$ in the range $6\cdot 10^{-7}<h<3\cdot
10^{-6}$. This gives $\psi=0.632\pm 0.021$.
The dashed line gives the $h^{-1/3}$ scaling law.

}

\bye